\documentclass[a4paper,10pt]{article}

\newcommand{\be}{\begin{equation}}
\newcommand{\ee}{\end{equation}}
\newcommand{\bea}{\begin{eqnarray}}
\newcommand{\eea}{\end{eqnarray}}

\newcommand{\nn}{\nonumber\\}

\begin{document}

\begin{center}

{\bf{\Large A new non-perturbative time-dependent string configuration}}

\vspace{0.5cm}

{\bf Jean Alexandre}

Department of Physics, King's College London, WC2R 2LS, UK

jean.alexandre@kcl.ac.uk

\vspace{1cm}

{\bf Abstract}

\end{center}

\vspace{0.5cm}

A time-dependent bosonic string configuration is discussed, in graviton and dilaton 
backgrounds, leading to Weyl-symmetry beta-functions which are
homogeneous in $X^0$, to any order in $\alpha^{'}$. As a consequence, 
a string reparametrization can always be implemented, such that beta functions can be cancelled,
to any order in $\alpha^{'}$.
This non-perturbative conformal invariance is valid for any target space dimension,
and leads to a power law expanding Universe, for which the power vanishes if a specific relation
between the dimension and dilaton amplitude holds.
Finally, $D=4$ is the minimum dimension (in the case of a spherical world sheet) for which this configuration
is consistent with a Wick rotation in a Minkowski target space.

\vspace{1cm}

The usual approach in String Cosmology consists in cancelling one-loop (or two-loop) beta functions corresponding to
Weyl invariance, and solving the corresponding differential equations in order to find the relevant
backgrounds. An alternative approach is discussed here \cite{AEM1}, where, instead of cancelling 
perturbative orders of the beta functions, we look for a configuration of the bosonic string 
leading to beta functions which are homogeneous functions of $X^0$, to all orders in $\alpha^{'}$.
This property will be enough to recover Weyl invariance, as the two-loop and higher orders in $\alpha^{'}$
of the beta functions are not fixed in a unique way, but depend on the renormalization scheme which 
is used when computing the trace of the energy-momentum tensor of the string \cite{metsaev}. Changing renormalization scheme 
corresponds to reparametrizing the string couplings, which in our case will not affect the time-dependence
of the configuration, but will only rescale the metric by a constant and add another constant to the dilaton.

In order to find a hint on what configuration can lead to homogeneous beta functions, 
we first present an exact functional method, dealing with the sigma model provided by the string 
action, where the evolution of the quantum theory with the amplitude of 
$\alpha^{'}$ is derived. We will see that, if the target space metric is conformally flat and proportional to
the second derivative of the dilaton (in the string frame), 
the latter is independent of the amplitude of quantum fluctuations.
Homogeneous beta functions will then be obtained for a specific case, where the graviton is a power law
of $X^0$.

\vspace{0.5cm}

We consider a bosonic string on a spherical world sheet, with bare action
\bea
S=\frac{1}{4\pi}\int d^2\xi\sqrt\gamma\left\lbrace 
\gamma^{ab}\lambda\eta_{\mu\nu}\partial_a X^\mu\partial_b X^\nu+R^{(2)}\phi_{bare}(X^0)\right\rbrace ,
\eea
where $\lambda=1/\alpha^{'}$ is left as a free parameter which controls the amplitude of 
quantum fluctuations. $\phi_{bare}(X^0)$ is a bare dilaton, whose precise 
form is not relevant, but it should in principle contain at least a cubic term in $X^0$, in order to generate 
quantum fluctuations. The quantum theory is obtained by defining the connected graphs generator functional
and its Legendre transform $\Gamma$, for a given value of $\lambda$. We obtain then a family of quantum
theories, labelled by $\lambda$ and described by $\Gamma_\lambda$. The exact evolution equation for
$\Gamma$ with $\lambda$ was derived in \cite{AEM1}, and leads to the following evolution for the dilaton
in the quantum theory (no summation on $i$: $g_{ii}$ is any of the space components of the metric):
\bea\label{evolphi}
g_{00}\dot\phi&=&-\frac{\Lambda^2}
{8\pi R^{(2)}}\left(1+(D-1)\frac{g_{00}}{g_{ii}}\right)\nn
&&+\frac{\phi^{''}}{4g_{00}}\ln\left(1+\frac{2\Lambda^2 g_{00}}{R^{(2)}\phi^{''}}\right),
\eea
where a dot represents a derivative with respect to $\lambda$ and a prime a derivative with respect to $X^0$.
$\Lambda$ is a {\it fixed} world sheet cut off, and we stress here the difference with a Wilsonian approach,
where $\lambda=1/\alpha^{'}$ would be fixed and the world sheet cut off would be running.
As a consistency check, one can easily see that a linear dilaton and flat metric
\cite{ABEN} lead to $\dot\phi=0$ (after
a rescaling of the space coordinates $X^i\to X^i\sqrt{D-1}$), which is expected, as such a configuration does not
generate quantum fluctuations. There is another possibility, thought, to obtain an $\alpha^{'}$-independent
dilaton: from the evolution equation (\ref{evolphi}), if $g_{00}$ is proportional to $g_{ii}$ and to
$\phi^{''}$, it is always possible to obtain $\dot\phi=0$ by rescaling the space coordinates $X^i$
(this rescaling absorbs the world sheet cut off). As a consequence, a necessary condition for the 
non-trivial solution to be independent of the amplitude of quantum fluctuations is that
\be\label{propto}
g_{\mu\nu}(X^0)\propto\phi^{''}(X^0)~\eta_{\mu\nu}.
\ee

\vspace{0.5cm}

We now turn to the conformal properties of a configuration satisfying the condition (\ref{propto}).
We first note that the one-loop beta functions corresponding to Weyl invariance cannot be cancelled by
a configuration satisfying the condition (\ref{propto}). But, for the following specific case 
\bea\label{config}
g_{\mu\nu}(X^0)&=&\frac{\kappa}{(X^0)^2}\eta_{\mu\nu}\nn
\phi(X^0)&=&\phi_0\ln X^0,
\eea
where $\kappa$ and $\phi_0$ are constants,
the different terms in the one-loop beta functions happen to be homogeneous to the same power of $X^0$. 
We checked that the 
next order in $\alpha^{'}$ again contains homogeneous functions of $X^0$, which are also
homogeneous to the first order. Finally, this property is valid to any order in $\alpha^{'}$: whatever power
of the Ricci or Riemann tensor is considered, and multiplied by covariant derivatives of the dilaton,  
contracting the indices with the metric or the inverse metric will always lead to the same power 
of $X^0$. As a consequence, we have for the configuration (\ref{config})
\bea
\beta^g_{00}&=&\frac{1}{(X^0)^2}\sum_{m=1}^\infty \xi_m\left(\frac{\alpha^{'}}{\kappa}\right)^m,\nn
\beta^g_{ij}&=&\frac{\delta_{ij}}{(X^0)^2}\sum_{m=1}^\infty \zeta_m\left(\frac{\alpha^{'}}{\kappa}\right)^m,\nn
\beta^\phi&=&\frac{1}{\alpha^{'}}\sum_{m=1}^\infty \eta_m\left(\frac{\alpha^{'}}{\kappa}\right)^m,
\eea
where $\xi_n,\zeta_n,\eta_n$ are coefficients independent of $\alpha^{'}$.\\
The next step is to argue that, from two-loops and above, the coefficients $\xi_n,\zeta_n,\eta_n$
are not unique but depend on the renormalization scheme which is used, in order to calculate 
the trace of the energy momentum tensor of the string \cite{metsaev}. A change of renormalization scheme
corresponds to a reparametrization of the string, which leaves the S matrix invariant, and reads,
at two loops
\bea\label{repara}
\tilde g_{\mu\nu}&=&g_{\mu\nu}+\alpha^{'}g_{\mu\nu}
\left(b_1R+b_2\partial^\rho\phi\partial_\rho\phi+b_3\nabla^2\phi\right),\nn
\tilde\phi&=&\phi+\alpha^{'}
\left(c_1R+c_2\partial^\rho\phi\partial_\rho\phi+c_3\nabla^2\phi\right),
\eea
where $b_1,...,c_1,...$ are any constants. In the specific case of the configuration (\ref{config}),
this reparametrization just rescales the metric by a constant factor and adds another constant to the dilaton.
But the important point is that the reparametrization (\ref{repara}) changes the beta functions in the following way \cite{metsaev}
\be\label{modif}
\tilde\beta^i=\beta^i+(\tilde g^j-g^j)\frac{\partial\beta^i}{\partial g^j}
-\beta^j\frac{\partial}{\partial g^j}(\tilde g^i-g^i),
\ee
such that it is always possible to choose the constants $b_1,...,c_1,...$ to cancel the two-loop
beta functions $ \tilde\beta^i$, what was done explicitly in \cite{AEM1}. We then conjecture that the
configuration (\ref{config}) satisfies Weyl invariance to any order in $\alpha^{'}$, since the 
reparametrization (\ref{repara}) can be extended to any order, and the modification
(\ref{modif}) is always valid. This cancellation obtained after a reparametrization
of the string is possible for any target space dimension, and for any dilaton amplitude.

\vspace{0.5cm}

Concerning Wilsonian properties of the configuration (\ref{config}), it was shown in \cite{AEM1} that
the latter is an infrared fixed point of momentum flows defined on the world sheet. To show this, an exact
renormalization equation was derived, following the approach given in \cite{WH}, where a sharp cut off was used (this is
indeed enough if one considers the evolution for the potential part only of the Wilsonian action).
As a consequence, the $\alpha^{'}$-fixed point solution of the equation (\ref{evolphi}) was 
identified with a world sheet Wilsonian IR fixed point.

\vspace{0.5cm}

In order to find the cosmological properties of the configuration (\ref{config}), we remind that the 
metrics in the string frame and the Einstein frame are related by
\be\label{identity}
dt^2-a^2(t)(d\vec x)^2
=\exp\left(\frac{-4\phi(x^0)}{D-2}\right)g_{\mu\nu}(x^0)dx^\mu dx^\nu,
\ee
where $t$ is the cosmic time and $a(t)$ is the scale factor of the corresponding spatially flat FRW Universe.
Plugging the configuration (\ref{config}) into the identity (\ref{identity}), we obtain the cosmic time as a 
power law of $x^0$, and the scale factor is
\be
a(t)= a_0t^{1+\frac{D-2}{2\phi_0}},
\ee
where $a_0$ is a constant. We obtain then a power-law expanding Universe, whose power vanishes if the
following relation holds
\be\label{agaga}
D-2+2\phi_0=0,
\ee
which is therefore the condition to have a Minkowski target space. Note that, if expressed in terms of the 
cosmic time, the dilaton is independent of $\phi_0$ and reads
\be
\phi=-\frac{D-2}{2}\ln t.
\ee

\vspace{0.5cm}

To conclude with the configuration (\ref{config}), we discuss its properties under a 
target space Wick rotation. Up to now, the world sheet metric had a Euclidean signature, but the
target space metric had a Minkowski signature.
Because of the logarithmic dilaton, the analytic continuation $X^0\to iX^0$
generates an imaginary part in the action.
As a consequence, the partition function acquires a phase and becomes, in the case of a spherical world sheet,
\be
Z\to Z_E\times\exp\left(i\pi\phi_0\right), 
\ee
where $Z_E$ is the Euclidean partition function (for both world sheet and target space metrics). 
In order to have a real partition function, we need 
$\phi_0$ to be an integer. Taking in addition the condition (\ref{agaga}) 
to have a Minkowski target space, we find that the allowed dimensions are all the even integers, starting from 4:
\be
D=4,6,8,\cdot\cdot\cdot
\ee
$D=4$ is thus the minimum dimension where the Wick rotation in target space is
consistent with the configuration (\ref{config}).

\vspace{0.5cm}

As a last remark, one can notice that the logarithmic dilaton, in $X^0$, leads to a diverging Euclidean
partition function, for late times, in the case $D=4$ and $\phi_0=-1$, for which the target space is flat and static.
This feature shows that a UV cut off in time is necessary for the consistency of the whole picture.
 
Finally, the extension of this work to a superstring would not change the results, as the homogeneity
of the beta functions corresponding to Weyl invariance would not be affected.


\begin{thebibliography}{99}


\bibitem{AEM1}
  J.~Alexandre, J.~Ellis and N.~E.~Mavromatos,
  JHEP {\bf 0612} (2006) 071
  [arXiv:hep-th/0610072].

\bibitem{metsaev}
  R.~R.~Metsaev and A.~A.~Tseytlin,
  Nucl.\ Phys.\  B {\bf 293} (1987) 385.

\bibitem{ABEN}
  I.~Antoniadis, C.~Bachas, J.~R.~Ellis and D.~V.~Nanopoulos,
  Nucl.\ Phys.\  B {\bf 328} (1989) 117.

\bibitem{WH}
  F.~J.~Wegner and A.~Houghton,
  Phys.\ Rev.\  A {\bf 8} (1973) 401.


\end{thebibliography}
\end{document}